\documentclass[a4paper, 11pt]{article}
\usepackage{jcappub}
\usepackage[T1]{fontenc}
\usepackage{amssymb, amsmath, amsfonts, amsbsy, graphicx, microtype, rotating, bm,
todonotes}
\usepackage{hyperref}
\usepackage{bm}
\usepackage{ulem}

\def\be{\begin{equation}}
\def\ee{\end{equation}}
\def\ba{\begin{eqnarray}}
\def\ea{\end{eqnarray}}

\usepackage{color}

\title{The effective field theory approach of teleparallel gravity, $f(T)$ gravity and 
beyond}

\author[a,b,c]{Chunlong Li,}
\author[d,e]{Yong Cai,}
\author[a,b,c]{\\ Yi-Fu Cai,}
\author[f,g,h]{Emmanuel N. Saridakis,}

\affiliation[a]{Department of Astronomy, University of Science and Technology of China, Hefei, Anhui 230026, China}
\affiliation[b]{CAS Key Laboratory for Researches in Galaxies and Cosmology, University of Science and Technology of China, Hefei, Anhui 230026, China}
\affiliation[c]{School of Astronomy and Space Science, University of Science and Technology of China, Hefei, Anhui 230026, China}
\affiliation[d]{School of Physics, University of Chinese Academy of Sciences, Beijing 100049, China}
\affiliation[e]{Department of Physics and Astronomy, University of Pennsylvania, Philadelphia, Pennsylvania 19104, USA}
\affiliation[f]{Department of Physics, National Technical University of Athens, Zografou Campus GR 157 73, Athens, Greece}
\affiliation[g]{CASPER, Physics Department, Baylor University, Waco, TX 76798-7310, USA}
\affiliation[h]{Chongqing University of Posts \& Telecommunications, Chongqing, 400065, China}

\emailAdd{chunlong@mail.ustc.edu.cn}
\emailAdd{caiyong13@mails.ucas.ac.cn}
\emailAdd{yifucai@ustc.edu.cn}
\emailAdd{Emmanuel\_Saridakis@baylor.edu}

\abstract{We develop the effective field theory approach to torsional modified gravities, a formalism that allows for the systematic investigation of the background and perturbation levels separately. Starting from the usual effective field theory approach to curvature-based gravity, we suitably generalize it at the background level by including terms of the contracted torsion tensor, and at the perturbation level by including pure torsion perturbative terms and mixed perturbative terms of torsion and curvature. Having constructed the  effective field theory action of general torsional modified gravity, amongst others we focus on $f(T)$ gravity and we perform a cosmological application. We investigate the scalar perturbations up to second order, and we derive the expressions of the Newtonian constant  and the post Newtonian parameter $\gamma$. Finally, we apply this procedure to two specific and viable $f(T)$ models, namely the power-law and the exponential ones, introducing a new parameter that quantifies the deviation from general relativity and depends on the model parameters. Since this parameter can be  expressed in terms of the scalar perturbation mode, a precise measurement of its evolution could be used as an alternative way to impose constraints on $f(T)$ gravity and break possible degeneracies between different $f(T)$ models.}


\begin{document}
	
%---------------------------------------
	
%\pacs{98.80.-k, 95.36.+x, 04.50.Kd, 04.30.Nk}

\maketitle
\flushbottom

\section{Introduction}

Current cosmological observations strongly indicate that the Universe underwent an early epoch of inflationary expansion, while it is undergoing a second accelerating epoch at present, which is difficult to be described by the combined paradigm of general relativity and standard model of particle physics \cite{Ade:2015xua, Joyce:2014kja}. In order to alleviate this difficulty one may either introduce the concepts of dark energy \cite{Copeland:2006wr, Cai:2009zp}, or modify the gravitational sector itself \cite{Nojiri:2006ri, Capozziello:2011et}.

In order to construct a gravitational modification one usually starts from general relativity and performs various extensions of the Einstein-Hilbert action, resulting for instance in $f(R)$ gravity \cite{DeFelice:2010aj}, $f(G)$ gravity \cite{Nojiri:2005jg}, etc. Nevertheless, one could equally well construct modified theories of gravity starting from the torsional gravitational formulation, namely from the Teleparallel Equivalent of General Relativity (TEGR) \cite{ein28,ein28b, Weitzen28, Hayashi79, Pereira:book, Maluf:2013gaa}. Since in this theory the Lagrangian $T$ is obtained by the contraction of the torsion tensor, one can construct various extensions, such as the $f(T)$ gravity \cite{Bengochea:2008gz, Linder:2010py}, $f(T_G)$ gravity \cite{Kofinas:2014owa, 
Kofinas:2014daa}, etc (see \cite{Cai:2015emx} for a comprehensive review). Despite the fact that general relativity and TEGR are equivalent at the level of equations, their modifications form different gravitational classes which need to be studied separately. 
Hence, torsional modified gravity has recently attracted the interest of the literature \cite{Ferraro:2006jd, Chen:2010va, Dent:2011zz, Wu:2010mn, Cai:2011tc, Li:2011rn, Bamba:2012vg, Wu:2011kh, Otalora:2013tba, Bahamonde:2015zma, Farrugia:2016qqe, Awad:2017tyz, Hohmann:2017jao,Bahamonde:2017wwk, Awad:2017yod, Hohmann:2018rwf, Abedi:2018lkr}.

In every gravitational theory, apart from the investigation at the background level one needs to proceed to the examination of the perturbations, since firstly these are related to the stability properties of the theory, and secondly, concerning cosmological 
backgrounds, the perturbations are straightforwardly related to various observables that in turn may be used to check and 
constrain the gravitational theory itself. In this direction a very helpful tool is the effective field theory (EFT) approach, which allows to ignore the details of the underlying fundamental theory and directly write a general action for the perturbations around a time-dependent background solution. Hence, one is able to systematically analyze the perturbations, separately from the background evolution. The EFT approach appeared first in \cite{ArkaniHamed:2003uy}, and it was then applied to inflation in \cite{Cheung:2007st} (see also \cite{ArkaniHamed:2007ky, Weinberg:2008hq, Ashoorioon:2018uey}) \footnote{The applications of this approach to nonsingular bounce cosmology were studied in \cite{Cai:2016thi, Cai:2017tku, Cai:2017dyi} with the associated observational 
implications being addressed in \cite{Cai:2014bea, Cai:2014zga, Cai:2014jla, Quintin:2015rta, Cai:2015vzv, Cai:2016hea, Li:2016xjb}.}. Additionally, the extension of this approach to dark energy was put forward in \cite{Creminelli:2008wc} and was further developed in \cite{Gubitosi:2012hu, Bloomfield:2012ff, Gleyzes:2013ooa, Bloomfield:2013efa, Piazza:2013coa}.

The structure of this paper is as follows. In Section \ref{Modelreview} we perform a review of teleparallel and $f(T)$ gravity. In Section \ref{TheEFTapproach} we present the EFT approach to torsional gravity in general, as well as of the sub-case of $f(T)$ 
gravity. Then in Section \ref{application} we perform an application of the EFT approach in a cosmological framework, both at the background and at the perturbation levels. In Section \ref{applicspecficmods} we apply the obtained formalism in two specific viable $f(T)$ models, that are known to pass the basic observational requirements, and we propose possible signatures of the $f(T)$ modification through post-Newtonian observations. Finally, Section \ref{Conclusion} is devoted to the 
conclusions with a discussion.

\section{Teleparallel and $f(T)$ gravity}
\label{Modelreview}

In this section we briefly review teleparallel and $f(T)$ gravity, and then we apply it in a cosmological framework. In torsional formulations of gravity ones uses the tetrad fields $e^{\mu}_A$ as the dynamical variable (Greek indices span the coordinate spacetime while Latin indices the tangent one), which form an orthonormal basis for the tangent space at each manifold point \cite{Cai:2015emx}. Additionally, we define the co-tetrad $e_{\mu}^A$ through $e^{\mu}_A e^A_{\nu} = \delta^{\mu}_{\nu}$ and $e^{\mu}_A e^B_{\mu} = \delta^A_B$. In order to handle orthonormality we introduce the tetrad metric $\eta_{AB} = \eta^{AB} = diag(-1, 1, 1, 1)$, and therefore the spacetime metric is given by
\begin{align}
\label{metric00}
 g_{\mu\nu}=\eta_{AB}e^A_{\mu}e^B_{\nu} ~.
\end{align}
Note that in this work we follow the mostly-plus signature \cite{Kofinas:2014daa} in order to be closer to standard gravity, instead of the mostly-minus one \cite{Bengochea:2008gz, Linder:2010py, Chen:2010va}. Nevertheless, the cosmological equations are the same in both conventions (although in some intermediate expressions there appear some sign differences).

In teleparallel gravity, concerning the connection ones uses the Weitzenb\"{o}ck one, namely $\hat{\Gamma}^{\lambda}_{\ \mu\nu} \equiv e^{\lambda}_A \partial_{\nu} e^A_{\mu} = - e^A_{\mu} \partial_{\nu} e^{\lambda}_{A}$, which leads to the torsion tensor
\begin{align}
\label{torntenr00}
 T^{\lambda}_{\ \mu\nu} \equiv \hat{\Gamma}^{\lambda}_{\ \nu\mu} - \hat{\Gamma}^{\lambda}_{\ \mu\nu}
 = e^{\lambda}_A(\partial_{\mu} e^A_{\nu} - \partial_{\nu} e^A_{\mu}) ~,
\end{align}
and consequently to the torsion scalar
\begin{equation}
\label{torsionsl00}
 T\equiv\frac{1}{4} T^{\rho \mu \nu} T_{\rho \mu \nu} + \frac{1}{2}T^{\rho \mu \nu }T_{\nu \mu \rho}
 - T_{\rho \mu }^{\ \ \rho }T_{\ \ \ \nu }^{\nu \mu} ~.
\end{equation}
The Weitzenb\"{o}ck connection is related to the Levi-Civita connection $\Gamma^{\sigma}_{\ \mu\nu}$ through
\begin{equation}
\label{LevCivWein}
 \hat{\Gamma}^{\rho}_{\ \mu\nu} = \Gamma^{\rho}_{\ \mu\nu} + {\cal{K}}^{\rho}_{\ \mu\nu} ~,
\end{equation}
where ${\cal{K}}^{\rho}_{\ \mu\nu}$ is the contorsion tensor defined as
\begin{align}
\label{Conten00}
 {\cal{K}}^{\rho}_{\ \mu\nu}\equiv\frac{1}{2}(T_{\mu\ \nu}^{\ \rho} + T_{\nu\ \mu}^{\ \rho} - T^{\rho}_{\ \mu\nu}) ~.
\end{align}
Hence, the covariant derivative of a quantity $A_{\mu}$ with respect to the Weitzenb\"{o}ck connection $\hat{\nabla}_{\nu}$ is related to its covariant derivative with respect to the Levi-Civita connection $\nabla_{\nu}$ through
\begin{equation}
\label{CovDerrel00}
 \hat{\nabla}_{\nu}A_{\mu}=\nabla_{\nu}A_{\mu}-{\cal{K}}^{\rho}_{\mu\nu}A_{\rho} ~.
\end{equation}
Since the  Riemann tensor corresponding to the Levi-Civita connection writes as
\begin{align}
 R^{\rho}_{\ \lambda\mu\nu} = \partial_{\mu} \Gamma^{\rho}_{\ \lambda\nu} + \Gamma^{\rho}_{\ \sigma\mu} \Gamma^{\sigma}_{\ \lambda\nu} - \partial_{\nu} \Gamma^{\rho}_{\ \lambda\mu} - \Gamma^{\rho}_{\ \sigma\nu} \Gamma^{\sigma}_{\ \lambda\mu} ~,
\end{align}
one can easily find that
\begin{align}
\label{RT}
 R = -T -2\nabla_{\mu} T^{\mu} ~,
\end{align}
where $T$ is the torsion scalar (\ref{torsionsl00}) corresponding to the Weitzenb\"{o}ck connection,  $T_{\mu}\equiv T^{\nu}{}_{\mu \nu}$ is the contraction of the torsion tensor, and $R$ is the Ricci scalar corresponding to the Levi-Civita connection.

In TEGR on uses the torsion scalar $T$ as the Lagrangian (similarly to the fact that one uses the Ricci scalar as the Lagrangian of general relativity), and variation with respect to the tetrads gives rise to exactly the same equations with general relativity. However, one can start from TEGR and construct gravitational modifications that do not have a corresponding curvature description. As a first such model, and inspired by the $f(R)$ extensions of general relativity, one can generalize $T$ to a function $f(T)$ and obtain $f(T)$ gravity, with action
\begin{align}
\label{fTaction}
 S = \int d^4x e\, \frac{M_P^2}{2}f(T) ~,
\end{align}
with $e = \text{det}(e_{\mu}^A) = \sqrt{-g}$ and where $M_P$ the Planck mass (we set the light speed to $c=1$). Variation of the above action, plus the matter part $S_m$, with respect to the tetrads provides the field equations as
\begin{align}
\label{fieldeqs}
 e^{-1}\partial_{\mu} (ee_A^{\rho}S_{\rho}{}^{\mu\nu}) f_{T} 
 + e_A^{\rho} S_{\rho}{}^{\mu\nu} \partial_{\mu}({T}) f_{TT} %\nonumber\\
 - f_{T} e_{A}^{\lambda} T^{\rho}{}_{\mu\lambda} S_{\rho}{}^{\nu\mu} + \frac{1}{4} e_{A}^{\nu} f({T}) = \frac{1}{2M_P^2} e_{A}^{\rho} \Theta_{\rho}{}^{\nu} ~,
\end{align}
with $f_{T}\equiv\partial f/\partial T$, $f_{TT}\equiv\partial^{2} f/\partial T^{2}$, with $\Theta_{\rho}{}^{\nu}$ the matter energy-momentum tensor corresponding to $S_m$, and where $S_\rho^{\:\:\:\mu\nu}$ is the super-potential defined as
\begin{eqnarray}
\label{superpot}
 S_\rho^{\:\:\:\mu\nu} \equiv \frac{1}{2} \Big( {\cal{K}}^{\mu\nu}_{\:\:\:\:\rho} 
 + \delta^\mu_\rho \: T^{\alpha\nu}_{\:\:\:\:\alpha} - \delta^\nu_\rho \: T^{\alpha\mu}_{\:\:\:\:\alpha} \Big) ~.
\end{eqnarray}

Let us now apply $f(T)$ gravity in a cosmological framework. We choose
\begin{equation}
\label{weproudlyuse}
 e_{\mu}^A={\rm diag}(1,a,a,a) ~,
\end{equation}
which corresponds to the flat Friedmann-Robertson-Walker (FRW) metric
\begin{equation}
\label{FRWmetric}
 ds^2= -dt^2+a^2(t)\,\delta_{ij} dx^i dx^j ~,
\end{equation}
with $a(t)$ the scale factor. One can immediately see that inserting the above tetrad into \eqref{torntenr00} and \eqref{torsionsl00} yields a useful relation between $T$ and the Hubble parameter $H=\frac{\dot a}{a}$, i.e., $T=6H^2$ (note that there is a sign difference comparing to some other works due to the convention of opposite metric signature). Furthermore, as usual we assume 
that the matter sector corresponds to a perfect fluid with energy density $\rho_m$ and pressure $p_m$. Thus, the general field equations (\ref{fieldeqs}) give rise to the two Friedmann equations
\begin{align}
\label{friedmann1}
 H^2 &= \frac{1}{3 M_{P}^2} \big( \rho_m +\rho_{DE}^{\text{eff}} \big) ~, \\
\label{friedmann2}
 \dot{H} &= -\frac{1}{2 M_{P}^2} \big( \rho_m +\rho_{DE}^{\text{eff}} +p_m +p_{DE}^{\text{eff}} \big) ~,
\end{align}
where we have defined
\begin{align}
\label{rhodeinit}
 \rho_{DE}^{\text{eff}} =& \frac{M_P^2}{2} \big( T -f +2T f_T \big) ~, \\
\label{pdeinit}
 p_{DE}^{\text{eff}} =& -\frac{M_P^2}{2} \Big[4\dot{H}(1+f_T+2Tf_{TT}) -f +T +2Tf_T \Big] ~.
\end{align}
The quantities $\rho_{DE}^{\text{eff}}$ and $p_{DE}^{\text{eff}}$ are respectively the effective dark energy density and pressure that arise from the $f(T)$ modification. Finally, note that the equations close by taking into account the conservations equations: 
$\dot{\rho}_m +3H (\rho_m+p_m)=0$ and $\dot{\rho}_{DE}^{\text{eff}} +3H (\rho_{DE}^{\text{eff}} +p_{DE}^{\text{eff}}) =0$.

\section{The EFT approach to torsional and $f(T)$  gravity}
\label{TheEFTapproach}

The main goal of the present work is to investigate torsional and $f(T)$ gravity through the EFT approach \cite{ArkaniHamed:2003uy}. By treating cosmological perturbations in a gravitational theory as the Goldstone modes of spontaneously broken time translations, one can construct the most general theory for the description of the fluctuations around a background evolution, attributing the investigation to the study of several typical operators. More precisely, the scalar perturbed degree of freedom can be ``eaten'' by the tetrad through a diffeomorphism transformation. Although this process breaks the full diffeomorphism invariance, it allows us to focus only on the geometrical quantities in order to formulate a more general theory (see, e.g. 
\cite{Piazza:2013coa}).

Before the detailed investigation, we make further comments on the symmetries of the theory under consideration. Besides the diffeomorphism invariance, we should also take into account the local Lorentz invariance, whose transformation writes as $e^{A}_{\mu} = \Lambda^A_B(x) e^{B}_{\mu}$ and $\eta^{CD} = \eta^{AB} \Lambda^C_A(x) \Lambda^D_B(x)$, since this issue has led to extensive discussions in the literature. One can see that a tetrad has $16$ independent components, of which only $10$ are needed to determine the metric tensor and hence can describe gravity, while six additional degrees of freedom are actually related to the possible ways of choosing the observer. In the usual formulation of teleparallel and $f(T)$ gravity, one imposes a zero spin connection and thus the only dynamical variable is the tetrad. Such a simplification, although useful in making the theory manageable in terms of solution extraction, has the disadvantage that the local Lorentz invariance is restricted \cite{Ferraro:2014owa}\footnote{Note that it is not manifest in the theory of teleparallel gravity in which field equations remain invariant under the local Lorentz transformations \cite{Obukhov:2006sk}, but becomes manifest in extended versions such as the $f(T)$ one.}. The approach to solving this problem is to construct a consistent and covariant formulation of $f(T)$ gravity, which is achieved if instead of the pure-tetrad formulation one uses both the vierbein and the spin connection in a way that for every vierbein choice a suitably constructed connection makes the whole theory covariant \cite{Krssak:2015oua}. In this case the $f(T)$ gravity can be frame-independent and Lorentz invariant.

Nevertheless, we point out that the above discussions on the Lorentz transformation in torsional modified gravities is not related to the EFT approach, which is only related to the diffeomorphism transformation. Hence, these two issues can be addressed separately. In particular, in the present study we focus on the EFT investigation.

\subsection{The basics of the EFT formalism}

In a theory of curvature-based gravity, a general action that arises from the EFT approach is given by \cite{Gubitosi:2012hu}:
\begin{align}
\label{curvEFTact}
 S &= \int d^4x \Big\{ \sqrt{-g} \Big[ \frac{M^2_P}{2} \Psi(t)R - \Lambda(t) - b(t)g^{00} \nonumber \\
 & \ \ \ \ \ \ \ \ \ \ \ \ \ \ \ \ \ \ \ \ \
 + M_2^4(\delta g^{00})^2 -\bar{m}^3_1 \delta g^{00} \delta K -\bar{M}^2_2 \delta K^2 - \bar{M}^2_3 \delta K^{\nu}_{\mu} \delta K^{\mu}_{\nu} \nonumber \\
 &\ \ \ \ \ \ \ \ \ \ \ \ \ \ \ \ \ \ \ \ \
 + m^2_2 h^{\mu\nu}\partial_{\mu} g^{00}\partial_{\nu}g^{00} +\lambda_1\delta R^2 + \lambda_2 \delta R_{\mu\nu}\delta R^{\mu\nu} +\mu^2_1 \delta g^{00} \delta R \Big] \nonumber \\
 & \ \ \ \ \ \ \ \ \ \ \ \ \
 + \gamma_1 C^{\mu\nu\rho\sigma} C_{\mu\nu\rho\sigma} + \gamma_2 \epsilon^{\mu\nu\rho\sigma} C_{\mu\nu}
^{\quad\kappa\lambda} C_{\rho\sigma\kappa\lambda} \nonumber \\
 & \ \ \ \ \ \ \ \ \ \ \ \ \
 +\sqrt{-g} \Big[ \frac{M^4_3}{3}(\delta g^{00})^3 -\bar{m}^3_2(\delta g^{00})^2 \delta K + ... \Big] \Big\} ~,
\end{align}
with $C^{\mu\nu\rho\sigma}$ the Weyl tensor, $\delta K^{\nu}_{\mu}$ the perturbation of the extrinsic curvature, and $R$ the Ricci scalar corresponding to the Levi-Civita connection. Moreover, we have introduced the time-dependent functions $\Psi(t)$, $\Lambda(t)$, $b(t)$, determined by the background evolution, and we have allowed for time-dependent coefficients in front of the 
various terms. We mention that, similarly to the gauge field theory with spontaneous symmetry breaking, we have imposed the unitary gauge, ``eating'' the would-be Goldstone bosons, and thus the theory displays only metric degrees of freedom.

The above action has been expanded around a given background solution, which is very similar to the expansion around a vacuum state in gauge theory. Thus, a significant advantage is that this process can organize the terms of the action as number of 
perturbations,  allowing one to separately deal with the background and the perturbations \cite{ArkaniHamed:2003uy}. The first line of action (\ref{curvEFTact}) determines the evolution of the background, the second, third and fourth lines are quadratic in perturbations, and therefore they affect the linear evolution of perturbations, while the last line is cubic in perturbations, and thus it affects the evolution of second-order perturbations.

\subsection{The EFT approach to torsional gravities}

We have now all the tools to proceed to the investigation of the  EFT approach to torsional gravities under the teleparallel condition. This procedure will lead to the inclusion of some new terms in the background and perturbation parts of action \eqref{curvEFTact}.

Firstly, we review how unitary gauge works. In a perturbed FRW geometry, the decomposition of a scalar degree of freedom reads as
\begin{equation}
 \phi(t,\vec{x}) = \phi_0(t) + \delta\phi(t,\vec{x}) ~.
\end{equation}
The unitary gauge corresponds to a choice of the time coordinate for which the perturbation of the scalar field vanishes. This implies that the coordinate $t$ is a function of $\phi$, i.e. $t=t(\phi)$,  and therefore $\delta\phi=0$ and thus the action displays metric degrees of freedom only. In the unitary gauge the theory is no longer invariant under four dimensional spacetime 
diffeomorphisms, but only  under the time dependent spatial diffeomorphisms (the unbroken symmetries), and this will be our guide to construct the action. Now, observing relations (\ref{LevCivWein}), (\ref{Conten00}) and (\ref{RT}), we deduce that we need to include both curvature and torsion terms. Thus, the EFT action can contain \cite{Piazza:2013coa}:
\begin{itemize}
\item i)
Terms that are invariant under all four-dimensional diffeomorphisms, such as the scalars $R$ and $T$, multiplied in general by functions of time.

\item ii)
Four dimensional covariant tensors with free zero upper indices, such as $g^{00}$, $R^{00}$ and $T^{0}$ ($T^{0}$ is the 0-index component of the contracted torsion tensor $T^{\mu}$).

\item iii)
Scalars constructed by spatial tensors. The spatial tensors include the spatial Riemann tensor $^{(3)}R_{\mu\nu\rho\sigma}$, the extrinsic curvature $K_{\mu\nu}$, the spatial torsion tensor $^{(3)}T^{\rho}_{\ \mu\nu}$, the ``extrinsic torsion'' $\hat{K}_{\mu\nu}$, etc.
\end{itemize}

The terms of type ii) and iii) arise from the definition of a preferred-time slicing by the scalar field $\phi$:
\begin{equation}	
 n_{\mu} = \frac{ \partial_{\mu} \phi(t) }{ \sqrt{-(\partial{\phi})^2}} = \frac{\delta^0_{\mu}}{\sqrt{-g^{00}}} ~.
\end{equation}
Due to the fact that the time-translation is broken, we can contract covariant tensors with this (orthogonal to the $t = {\rm const}$ surfaces) unitary vector, resulting to terms with upper $0$ indices. Considering the Weitzenb\"ock covariant derivative of $n_{\mu}$, and projecting it on the surface of constant $t$, we have
\begin{equation}
 h^{\sigma}_{\mu} \hat{\nabla}_{\sigma} n_{\nu} \equiv \hat{K}_{\mu\nu} ~,
\end{equation}
with $h_{\mu\nu} \equiv g_{\mu\nu} +n_{\mu\nu}$ the induced metric of this surface. We can verify that this quantity is a spatial tensor, and we refer to it as ``extrinsic torsion''. Recalling the relation (\ref{CovDerrel00}) between the ordinary covariant derivative and the Weitzenb\"ock covariant derivative, we can find
\begin{align}
\label{Kmunu00}
 \hat{K}_{\mu\nu} \equiv h^{\sigma}_{\mu} \hat{\nabla}_{\sigma} n_{\nu} 
 = K_{\mu\nu} - {\cal K}^{\lambda}_{\ \nu\mu} n_{\lambda} + n_{\mu} \frac{1}{g^{00}} T^{00}_{\quad\nu} ~,
\end{align}
which is the relation between extrinsic curvature and extrinsic torsion. Contracting its two indices we obtain
\begin{align}
\label{K00}
 K = & \nabla_{\mu}n^{\mu} = \hat{\nabla}_{\mu}n^{\mu} -{\cal K}^{\mu}_{\ \lambda\mu}n^{\lambda} \nonumber \\
 = & \hat{K} +T_{\lambda} n^{\lambda} = \hat{K} + (-g^{00})^{-1/2} T^0 ~.
\end{align}
In summary, knowing \eqref{Kmunu00} and \eqref{K00} and observing the action \eqref{curvEFTact}, it is implied that if we include $T^{00\nu}$ and $T^0$ in the action, then we can avoid the use of $\hat{K}$ and $\hat{K}_{\mu\nu}$.

Let us now consider the covariant derivative of $n_{\mu}$ perpendicular to the time slicing, namely
\begin{equation}	
 n^{\sigma} \hat{\nabla}_{\sigma} n_{\nu} = n^{\sigma} \nabla_{\sigma} n_{\nu} + \frac{1}{g^{00}}T^{00}_{\quad \nu} ~.
\end{equation}
The term $n^{\sigma} \nabla_{\sigma} n_{\nu}$ leads to a term containing $g^{00}$ and $h^{\mu}_{\nu}$ \cite{Cheung:2007st}, which has already been included in action (\ref{curvEFTact}). Hence, we deduce that if we use $T^{00\nu}$ in the action then we can also avoid $n^{\sigma} \hat{\nabla}_{\sigma} n_{\nu}$.

We proceed by constructing action terms corresponding to spatial diffeomorphisms invariant scalars, arising from three-dimensional spatial tensors. In order to do this we use the unit normal vector $n_{\mu}$, or the projection operator $h^{\mu}_{\nu}$. For convenience we use only four-dimensional tensors, since their three-dimensional counterparts can always arise from the normal vector or the projection operator, as for example
\begin{align}
 ^{(3)}R_{\mu\nu\rho\sigma} = h^{\alpha}_{\mu} h^{\beta}_{\nu} h^{\gamma}_{\rho} h^{\delta}_{\sigma}
 R_{\alpha\beta\gamma\delta} - K_{\mu\rho} K_{\nu\sigma} + K_{\nu\rho} K_{\mu\sigma} ~,
\end{align}
and thus they can be used in our action. Finally, the use of spatial covariant derivatives can also be avoided, due to the fact that the spatial covariant derivative of a spatial tensor can be extracted from the projection of the four-dimensional covariant derivative.

Let us now examine the spatial torsion tensor $^{(3)}T^{\rho}{}_{\mu\nu}$. We use $D_a$ to represent the Weitzenb\"ock covariant derivative belonging to the constant-time surface, defined as
\begin{align}\label{D definition}
 D_cT^{a_1\cdots a_k}{}_{b_1\cdots b_k} \equiv  h^{a_1}{}_{d_1}\cdots h^{a_k}{}_{d_k}h_{b_1}{}^{e_1}
 \cdots h_{b_l}{}^{e_l} \cdot h_c{}^f\hat{\nabla}_f T^{d_1\cdots d_k}{}_{e_1\cdots e_l} ~.
\end{align}
The presence of torsion in the constant-time surface implies that the covariant derivatives of a scalar field $\phi$ do not commute, since
\begin{align}
\label{surface D}
 (D_aD_b-D_bD_a)\phi=-^{(3)}T^c{}_{ab}D_c\phi ~.
\end{align}
Similarly, for the four dimensional spacetime we have
\begin{align}  	
\label{spacetime D}
 (\hat{\nabla}_a\hat{\nabla}_b -\hat{\nabla}_b\hat{\nabla}_a)\phi = - ^{(3)}T^c{}_{ab} \hat{\nabla}_c \phi ~.
\end{align}
Using \eqref{D definition}, \eqref{surface D} and \eqref{spacetime D}, we can derive that the spatial torsion tensor is just the projection of the full spacetime torsion tensor, i.e.,
\begin{align}
 h^d_ah^c_bh^f_eT^e{}_{dc}= ^{(3)}\!T^f{}_{ab} ~.
\end{align}
In summary, we can use $T^e{}_{dc}$ or $^{(3)}T^f{}_{ab}$ to be present in our action.

As a next step we arrange all the above operators in powers of number of perturbations, expanding around an FRW background. Let us consider an operator made by the contraction of two tensors $X$ and $Y$ (we can straightforwardly generalize it to more tensors). Expanding them linearly as $X\approx X^{(0)}+\delta X$ and $Y\approx Y^{(0)}+\delta Y$, we have
\begin{align}
\label{TGexpr}
 XY = \delta X \delta Y + X^{(0)}Y + XY^{(0)} - X^{(0)} Y^{(0)} ~.
\end{align}
Note that in an FRW background one can always express the unperturbed tensors $X^{(0)}$ and $Y^{(0)} $ as functions of $g_{\mu\nu}$, $n_{\mu}$ and $t$. Thus, concerning the Riemann and torsion tensors as well as the extrinsic curvature and torsion, we can express them in turn as follows:
\begin{align}
\label{Rexpres001}
 R_{\mu\nu\rho\sigma}^{(0)} =& f_{1}(t) g_{\mu\rho} g_{\nu\sigma} + f_2(t) g_{\mu\rho}n_{\nu} n_{\sigma} + f_3(t) g_{\mu\sigma} g_{\nu\rho} \nonumber \\
 & +f_4(t) g_{\mu\sigma} n_{\nu} n_{\rho} + f_5(t) g_{\nu\sigma} n_{\mu} n_{\rho} +f_6(t) g_{\nu\rho} n_{\mu} n_{\sigma} ~, \\
\label{Rexpres002}
 T^{(0)}_{\rho\mu\nu} =& g_1(t) g_{\rho\nu} n_{\mu} + g_2(t) g_{\rho\mu} n_{\nu} ~, \\
\label{Rexpres003}
 K^{(0)}_{\mu\nu} =& f_7(t) g_{\mu\nu} + f_8(t) n_{\mu} n_{\nu} ~, \\
\label{Rexpres004}
 \hat{K}^{(0)}_{\mu\nu} =& 0 ~,
\end{align}
where the various time-dependent terms are determined by the background evolution. Additionally, we mention that the expressions (\ref{Rexpres001})-(\ref{Rexpres004}) hold modulo a factor of polynomials of $g^{00}$, which is irrelevant for our analysis, since the term $X^{(0)}Y^{(0)}$ in (\ref{TGexpr}) is just a polynomial of $g^{00}$ with time-dependent 
coefficients.

The first term in \eqref{TGexpr} starts  quadratic in perturbations and therefore we keep it. In the second term, i.e., $X^{(0)}Y$, by construction $Y$ is linear in $R_{\mu\nu\rho\sigma}$, $K_{\mu\nu}$,  $T^{\rho}_{\ \mu\nu}$ and $\hat{K}_{\mu\nu}$, with covariant derivatives acting on them. Having in mind the relation of the two connections \eqref{LevCivWein}, it is sufficient to consider 
only the covariant derivative with respect to the Levi-Civita connection, which can then be handled using successive integrations by parts acting on $X^{(0)}$ and the time-dependent coefficient. Thus, this procedure will produce extrinsic curvature terms. After contracting all the indices, we will get the only possible scalar linear terms with no covariant derivatives, namely $K$, $R^{00}$, $R$, $\hat{K}$, $T^0$ and $T$, and due to relation (\ref{K00}) we can eliminate $\hat{K}$ in terms of $T^0$. As it was shown in \cite{Cheung:2007st}, the integration of $R^{00}$ and $K$ with time-dependent coefficients gives just the linear operator $g^{00}$ along with invariant terms that start quadratically in the perturbations, and hence we deduce that these two terms can also be avoided in the background action. Lastly, using (\ref{RT}) it is implied that the integration of the boundary term with a time-dependence coefficient becomes
\begin{align}
 \int d^4x \sqrt{-g} h(t) \nabla_{\mu} T^{\mu} = -\int d^4x \sqrt{-g} \dot{h}(t)T^0 ~.
\end{align}

We now have all the material to assemble the EFT action of torsional gravity. Since the remaining background terms are $R$ and $T^0$, we result to
\begin{align}
\label{actionfin}
 S = \int d^4x \sqrt{-g} \Big[ \frac{M^2_P}{2} \Psi(t)R - \Lambda(t) - b(t) g^{00} + \frac{M^2_P}{2} d(t) T^0 \Big] + S^{(2)} ~,
\end{align}
with $d(t)$ a time-dependent function. In the above action we have included the part $S^{(2)}$, which includes all terms that explicitly start quadratic in perturbations, and thus it does not affect the background dynamics. In addition to the terms presented in action \eqref{curvEFTact}, the part $S^{(2)}$ is expected to include also:\\
i) Pure torsion terms such as $\delta T^2$, $\delta T^0\delta T^0$ and $\delta T^{\rho\mu\nu} \delta T_{\rho\mu\nu}$ (since in the action we include $T^{0}$, due to (\ref{K00}) we can avoid the presence of $\hat{K}$);\\
ii) Terms that mix curvature and torsion, such as $\delta T\delta R$, $\delta g^{00}\delta T$, $\delta g^{00}\delta T^0$ and $\delta K\delta T^0$.

In principle, the action (\ref{actionfin}) could include any modified theories based on teleparallel gravity. However, since up to now we do not have a most general modified teleparallel gravity (in analogy with Horndeski theory as one of the most general form in 
the field of curvature-based scalar tensor theory), we shall not specify the full form of $S^{(2)}$. Instead, we wish that the EFT approach developed in the present study could shed light on the future investigations of this subject. In the following, we will use 
the $f(T)$ gravity as an example to illustrate how to apply the EFT approach to a specific theory.

\subsection{The EFT approach to $f(T)$ gravity}
\label{The EFT form of $f(T)$ theory}

In the previous subsection we have investigated the EFT approach to torsional gravity under the teleparallel condition. Thus, we can now apply the procedure to $f(T)$ gravity. A first issue that need to be handled is how to incorporate the time slicing. Similar to 
the EFT approach to the $f(R)$ gravitational paradigm \cite{Gubitosi:2012hu}, we start by expanding the action (\ref{fTaction}) as
\begin{align}
 S = \frac{M^2_P}{2} \int d^4x \sqrt{-g} \Big[ f_{T}(T^{(0)})T + f(T^{(0)}) - f_T(T^{(0)})T^{(0)} +\frac12 f_{TT}(T^{(0)})\delta T^2 + ... 
\Big] ~,
\end{align}
where $T^{(0)}$ is the torsion scalar at the background order. Then, we fix the time slicing in order to coincide with the uniform $T$ hypersurfaces, since doing so would make the terms in the above expansion beyond the linear order to vanish due to the fact that 
their contribution to the equations of motion would always include at least one power of $\delta T$.

As a result, at the background level, we construct the unitary-gauge action to be
\begin{align}
\label{action}
 S = \frac{M^2_{P}}{2} \int d^4x \sqrt{-g} \Big[ -f_T(T^{(0)})R +2\dot{f_T}(T^{(0)}) T^0 - T^{(0)} f_T (T^{(0)}) + f(T^{(0)}) \Big] ~.
\end{align}
By observing this expression we deduce that it ought to be of the form (\ref{actionfin}), with
\begin{align}
\label{fTidentification}
 \Psi(t) &= -f_T(T^{(0)}) ~, \nonumber \\
 \Lambda(t) &= \frac{M^2_{P}}{2}\left[T^{(0)}f_T(T^{(0)})-f(T^{(0)})\right]  ~, \nonumber \\
 d(t) &= 2\dot{f}_T(T^{(0)}) ~,\nonumber \\
 b(t) &= 0 ~.
\end{align}
Finally, if the $T^0$ term vanishes, the above action can reproduce the usual EFT form of $f(R)$ gravity provided in \cite{Gubitosi:2012hu}.

One interesting observation is the following: although the action of $f(T)$ gravity in general cannot be decomposed as $R$ plus a boundary term, similar to the case of TEGR, using the EFT approach we are able to write the action as $R$ plus a ``boundary term'', and hence study the features and the properties  of $f(T)$ theory by studying the ``boundary term'' $T^0$. Furthermore, given that $T^0$ is only the contraction of the torsion tensor $T^{\lambda}{}_{\mu\nu}$, while $T$ is quadratic with respect to 
$T^{\lambda}{}_{\mu\nu}$, we deduce that the EFT approach can simplify significantly the involved calculations, and moreover it allows to use information obtained from the well-studied EFT approach of curvature modified gravity.

We close this subsection by mentioning that, since in the previous subsection we presented the EFT approach to a general torsional gravity under the teleparallel condition, apart from $f(T)$ gravity we may construct the EFT action of other torsional modified gravity subclasses as well, by choosing sub-cases of the action (\ref{actionfin}). For instance, one may study the scalar-torsion theories \cite{Geng:2011aj} in which the terms $f(\phi)T$ or $\partial_{\mu}\phi\partial^{\mu}\phi T$ \cite{Kofinas:2015hla, Abedi:2017jqx} can be incorporated through the unitary gauge, letting $\phi=\phi(t)$, and thus contribute to $S^{(2)}$ with terms like $\delta g^{00}\delta T$. Additionally, the $f(T,R)$ formalism \cite{Myrzakulov:2012qp}, with a given Taylor expansion around the 
background $T^{(0)}$ and $R^{(0)}$ being expressed as
\begin{align}
 f(T,R) =& f(T^{(0)},R^{(0)}) + \frac{\partial f(T^{(0)},R^{(0)})}{\partial R}(R - R^{(0)}) + \frac{\partial f(T^{(0)},R^{(0)})}{\partial T}(T -T^{(0)}) \nonumber \\
 & +\frac{1}{2} \frac{\partial^2 f(T^{(0)}, R^{(0)})}{\partial R^2} \delta R^2 + \frac{\partial^2 f(T^{(0)}, R^{(0)})}{\partial R \partial T}\delta T \delta R + \frac{1}{2} \frac{\partial^2 f(T^{(0)},R^{(0)})} {\partial T^2} \delta T^2 + ... ~,
\end{align}
can also be straightforwardly studied under the EFT-approach action (\ref{actionfin}).

\section{Application to Cosmology}
\label{application}

In the previous section we presented the EFT approach to torsional and $f(T)$ gravity. Hence, we have all the machinery needed in order to proceed to the application of the EFT approach in a cosmological framework. In the following two subsections we perform this application at the background and at the perturbation levels.

\subsection{The background evolution}
\label{The background evolution}

As we have mentioned previously, in order to study the cosmological application of $f(T)$ gravity, we consider as usual the flat FRW metric \eqref{FRWmetric}, which arises from the FRW tetrad choice \eqref{weproudlyuse}. Inserting these expressions into the first line of the action \eqref{actionfin}, with the addition of the matter part $S_m$, and performing variation we obtain
\begin{align}
\label{cc}
 b(t) &= M_{P}^2 \Psi \Big( -\dot{H} - \frac{\ddot{\Psi}}{2\Psi} + \frac{H\dot{\Psi}}{2\Psi} - \frac{\dot{d}}{4\Psi} + \frac{3Hd}{4\Psi} \Big) -\frac{1}{2} \big( \rho_m+p_m \big) ~, \\
\label{ll}
 \Lambda(t) &= M_{P}^2\Psi \Big( 3H^2 + \frac{5H \dot{\Psi}}{2\Psi} + \dot{H} + \frac{\ddot{\Psi}}{2\Psi} + \frac{\dot{d}}{4\Psi} +\frac{3Hd}{4\Psi} \Big) -\frac{1}{2} \big( \rho_m - p_m \big) ~,
\end{align}
which are just the two Friedmann equations of $f(T)$ gravity. As usual, by focusing on the post-inflationary eras, we can re-write the above Friedmann equations to their standard form, namely
\begin{align}
\label{f11}
 H^2 &= \frac{1}{3 M_{P}^2} \big( \rho_m +\rho_{DE}^{\text{eff}} \big) ~, \\
\label{f22}
 \dot{H} &= -\frac{1}{2 M_{P}^2} \big( \rho_m +\rho_{DE}^{\text{eff}} +p_m +p_{DE}^{\text{eff}} \big) ~,
\end{align}
where
\begin{align}
\label{rhoDE1}
 \rho_{DE}^{\text{eff}} =& b+\Lambda -3 M_P^2 \Big[ H\dot{\Psi}+\frac{dH}{2}+H^2(\Psi-1) \Big] ~, \\
\label{pDE1}
 p_{DE}^{\text{eff}} =& b-\Lambda +M_P^2 \Big[ \ddot{\Psi} +2H\dot{\Psi} + \frac{\dot{d}}{2} + (H^2+2\dot{H})(\Psi-1) \Big] ~.
\end{align}
The quantities $\rho_{DE}^{\text{eff}}$ and $p_{DE}^{\text{eff}}$ are respectively the effective dark energy density and pressure arising from the EFT approach in the general torsional gravity. In the case of $f(T)$ gravity, as we showed in the previous section, one needs to make the identification (\ref{fTidentification}) in the general EFT action. Hence, if we insert (\ref{fTidentification}) into (\ref{rhoDE1}) and (\ref{pDE1}), we finally obtain the effective dark energy density and pressure in the case of $f(T)$ gravity, as they arise from the EFT approach, namely
\begin{align}
\label{rhodein22}
 \rho_{DE}^{\text{eff}} =& \frac{M_P^2}{2} \Big[ T^{(0)}-f(T^{(0)})+2T^{(0)}f_T(T^{(0)}) \Big] ~, \\
 p_{DE}^{\text{eff}} =& -\frac{M_P^2}{2} \Big[ 4\dot{H} \big[ 1+f_T(T^{(0)}) + 2T^{(0)}f_{TT}(T^{(0)} ) \big] - f(T^{(0)}) +T^{(0)} +2T^{(0)} f_T(T^{(0)}) \Big] ~,
 \label{pdein22}
\end{align}
where $T^{(0)}=6H^2$ is the torsion scalar at the background level. These expressions are exactly in agreement with \eqref{rhodeinit} and \eqref{pdeinit}. Thus, through the EFT approach we are able to re-obtain the $f(T)$ cosmological equations at the background level.

\subsection{Scalar perturbations of $f(T)$ gravity through the EFT approach}
\label{scalarpert}

In the previous subsection we extracted the equations of $f(T)$ cosmology at the background level, following the general EFT approach to torsional gravity. In particular, we started from the general action (\ref{actionfin}), we set the perturbation part $S^{(2)}$ to zero, and then we performed our analysis imposing in the end the $f(T)$ identification (\ref{fTidentification}). In the present subsection we are interested in studying the perturbations of $f(T)$ gravity around a cosmological background, through the EFT approach. For simplicity we will focus on scalar perturbations, while the investigation of the tensor perturbations was performed in \cite{Cai:2018rzd}.

Taking the degrees of freedom released by local Lorentz violation into consideration, the perturbed tetrads have the form of \cite{Izumi:2012qj}
\begin{align}
 e^0_{\mu}&=\delta^0_{\mu}+\delta^0_{\mu}\phi+a\delta^i_{\mu}(\partial_i \chi+\chi_i) ~, 
\nonumber \\
 e^a_{\mu}&=a\delta^i_{\mu}\delta^a_i+\delta^0_{\mu}\delta^a_i(\partial^i{\cal E}+{\cal E}^i) \nonumber \\
 &+a\delta^i_{\mu}\delta^a_j\Big[\epsilon_{ijk}(\partial_k\sigma+V_k)-\psi\delta_{ij} +\frac12(h_{ij} + \partial_i\partial_j F + \partial_j C_i+\partial_i C_j)\Big] ~,
\label{perturbation of tetrads}
\end{align}
where the small Latin indices represent 1,2,3. Eqs. (\ref{perturbation of tetrads}) gives rise to the usual perturbed metric
\begin{align}
 g_{00} &= -1 -2\phi ~, \label{g00} \\
 g_{0i} &= a(\partial_i B+B_i) ~, \label{g0i} \\
 g_{ij} &= a^2 [(1-2\psi)\delta_{ij} +\partial_i\partial_j F +\partial_j C_i +\partial_i C_j +h_{ij} ] ~,
\label{perturbed metric}
\end{align}
where $B={\cal E}-\chi$, $B_i={\cal E}_i-\chi_i$. We note that for tetrads there are six more perturbed degrees of freedom than the metric, i.e. the scalar $\chi$ (or ${\cal E}$), the transverse vector $\chi_i$ (or ${\cal E}_i$), the pseudoscalar $\sigma$ and the 
pseudovector $V_k$. As was pointed out in \cite{Wu:2012hs}, they are the degrees of freedom released by the local Lorentz violation.

From now on we only focus on the scalar perturbations:
\begin{align}
\label{g00} g_{00} &= -1 -2\phi ~, \\
\label{g0i} g_{0i} &= a\partial_i B ~, \\
\label{perturbed metric} g_{ij} &= a^2 [(1-2\psi)\delta_{ij} +\partial_i\partial_j F ] ~,
\end{align}
and the corresponding perturbed tetrads are
\begin{align}
 e^0_{\mu} =& \delta^0_{\mu} +\delta^0_{\mu}\phi +a\delta^i_{\mu} \partial_i\chi ~, \\
 e^a_{\mu} =& a\delta^i_{\mu}\delta^a_i + \delta^0_{\mu} \delta^a_i \partial^i{\cal E} + a\delta^i_{\mu}\delta^a_j \big[ \epsilon_{ijk} \partial_k \sigma -\psi\delta_{ij} + \frac{1}{2} \partial_i \partial_j F \big] ~.
\end{align}

The gauge transformation reads as
\begin{align}
 \delta\tilde{e}^a_{\mu} = \delta e^a_{\mu} -\xi^{\alpha} \bar{e}^a_{\mu,\alpha} - \xi^{\rho}_{,\mu} \bar{e}^a_{\rho} ~,
\end{align}
which leads to the gauge transformation of scalar perturbations as
\begin{align}
 & \tilde{\phi} = \phi-\xi^0,_{0} ~,~ \tilde{\chi} = \chi-\frac{1}{a}\xi^0 ~,~ \tilde{{\cal E}} = {\cal E}-\xi_{,0} a ~, \nonumber\\
 & \tilde{\psi} =\psi-\xi^0\frac{\dot{a}}{a} ~,~ \tilde{F} = F-2\xi ~,~ \tilde{\sigma} = \sigma ~,
\end{align}
where we have decomposed $\xi^i$ as $\xi^i=a^{-1}(\partial_i \xi+\xi^{tr}_i)$, with $\xi^{tr}$ referring to the transverse part of $\xi^i$. Thus, the gauge transformation of scalars depends on two transformation parameters and we can set two of these scalars equal to zero. For convenience we choose the familiar Newtonian gauge, setting $B={\cal E}-\chi=0$ and $F=0$, resulting to
\begin{align}
 e^0_{\mu} &= \delta^0_{\mu} +\delta^0_{\mu}\phi +a\delta^i_{\mu}\partial_i\chi ~,  \\
 e^a_{\mu} &= a\delta^i_{\mu}\delta ^a_i +\delta^0_{\mu} \delta^a_i \partial^i \chi + a\delta^i_{\mu} \delta^a_j \big( \epsilon_{ijk}\partial_k\sigma-\delta_{ij}\psi \big) ~.
\end{align}

If the theory preserves parity, the pseudoscalar $\sigma$ cannot be present in the first-order scalar perturbation equations, and therefore the perturbed tetrads are given by
\begin{align}
 e^0_{\mu} &= \delta^0_{\mu} +\delta^0_{\mu}\phi +a\delta^i_{\mu} \partial_i\chi ~,  \\
 e^a_{\mu} &= a\delta^i_{\mu} \delta^a_i(1-\psi) +\delta^0_{\mu}\delta^a_i \partial^i\chi ~.
\end{align}
Here we note that there is an additional scalar $\chi$ present in the perturbation theory and we will study its effect later on. In a given coordinate system, we can always write a tetrad field $e^a_{\mu}$ as $\delta^a_{\nu}e^{\nu}{}_{\mu}$, and thus use the exponent expansion 
$e^{\nu}{}_{\mu} =e^{\delta e^{\nu}{}_{\mu}} =\delta^{\nu}{}_{\mu} +\delta e^{\nu}{}_{\mu} +1/2(\delta e^{\nu}{}_{\mu})^2 +...$, 
where $\delta e^{\nu}{}_{\mu}$ is the perturbation of $e^{\nu}{}_{\mu}$. Then we can obtain the perturbed tetrads up to second order, which are expressed as follows:
\begin{align}   	
 e^0_{\mu} =& \delta^0_{\mu} \big( 1 +\phi +\frac12\phi^2 +\frac12\partial_i\chi \partial_i\chi \big) + a\delta^i_{\mu} \Big[ \partial_i \chi + \frac12(\phi\partial_i\chi -\psi\partial_i\chi) \Big] ~, \\
 e^a_{\mu} =& a\delta^i_{\mu} \delta^a_i \big( 1-\psi+\frac12\psi^2 \big) + \frac{a}{2} \delta^i_{\mu} \delta^a_j \partial_i \chi \partial_j \chi + \delta^0_{\mu} \delta^a_i \Big[ \partial_i\chi +\frac12(\phi\partial_i\chi - \psi\partial_i\chi) \Big] ~, \\
 e^{\mu}_0 =& \delta^{\mu}_0 \big( 1 -\phi +\frac12\phi^2 +\frac12\partial_i \chi\partial_i\chi \big) + \frac1a\delta_i^{\mu} \Big[ -\partial_i\chi +\frac12(\phi\partial_i\chi - \psi\partial_i\chi) \Big] ~, \\
 e^{\mu}_a =& \frac1a \delta_i^{\mu} \delta_a^i \big( 1 + \psi + \frac12 \psi^2 \big) + \frac{1}{2a} \delta_i^{\mu} \delta_a^j \partial_i \chi \partial_j \chi + \delta_0^{\mu} \delta_a^i \Big[ -\partial_i\chi +\frac12(\phi\partial_i\chi - \psi \partial_i \chi) \Big] ~.
\end{align}
Accordingly, one can also write down the perturbed metric up to second order, namely
\begin{align}
 g_{00} &= -(1+2\phi+2\phi^2) ~, \\
 g_{ij} &= a^2\delta_{ij}(1-2\psi+2\psi^2) ~, \\
 g^{00} &= -(1-2\phi+2\phi^2) ~, \\
 g^{ij} &= a^{-2}\delta_{ij}(1+2\psi+2\psi^2) ~,
\end{align}
as well as
\begin{align}
\label{g}
 \sqrt{-g} = a^3 (1 -3\psi +\phi +\frac{9}{2}\psi^2 -3\phi\psi +\frac12\phi^2) ~.
\end{align}

Up to now we have described all the geometrical ingredients of the perturbation incorporation. Hence, we are ready to proceed to the examination of the perturbation in the gravitational theory at hand.

A first issue to be handled is that the action (\ref{action}) was written in the unitary gauge, while the above perturbation description was performed in the Newtonian gauge. In order to connect the two gauge descriptions we need to apply the St\"uckelberg trick to 
introduce the Goldstone modes and restore diffeomorphism invariance. We start by calculating $T^0$ up to second order as
\begin{align}
\label{T0relat}
 T^0 = g^{0\mu}T^{\nu}_{\ \mu\nu} =& -3H +6H\phi +3\dot{\psi} -6H\phi^2 -6\dot{\psi}\phi \nonumber \\
 & + \frac{1}{a} \partial_i \partial_i \chi - \frac{1}{2a} \partial_i \phi \partial_i \chi - \frac{3}{2a} \phi \partial_i \partial_i \chi -\frac{1}{2a}\partial_i \psi \partial_i \chi + \frac{1}{2a} \psi \partial_i \partial_i \chi ~,
\end{align}
and $T^i$ up to first order as
\begin{align}
 T^i=a^{-2}(\partial_i\phi-2\partial_i\psi)-\frac{1}{a}\partial_i\dot{\chi} ~.
\end{align}
The St\"uckelberg trick can be expressed as
\begin{align}
\label{Stueckelberg trick 1}
 \dot{f}_T \longrightarrow \dot{f}_T +\ddot{f}_T\pi +\frac{1}{2}\dddot{f}_T\pi^2 + ... ~,
\end{align}
and thus relation (\ref{T0relat}) then becomes
\begin{align}
\label{Stueckelberg trick 2}
 T^0 \longrightarrow & T^0+\partial_{\mu}\pi T^{\mu} \nonumber \\
 = & -3H + 6H\phi -6H\phi^2 +3\dot{\psi} -6\dot{\psi}\phi + \dot{\pi}(-3H+6H\phi+3\dot{\psi}) \nonumber \\
 & +a^{-2}\partial_i\pi(\partial_i\phi-2\partial_i\psi) + \frac{1}{a}\partial_i\partial_i\chi -\frac{1}{2a}\partial_i\phi\partial_i\chi 
-\frac{3}{2a}\phi\partial_i \partial_i\chi \nonumber\\
 & -\frac{1}{2a}\partial_i\psi\partial_i\chi +\frac{1}{2a}\psi\partial_i\partial_i\chi + \frac{1}{a}\dot{\pi}\partial_i\partial_i\chi 
-\frac{1}{a}\partial_i\pi\partial_i\dot{\chi} ~.
\end{align}

In principle we can follow a similar method for all terms of the action (\ref{action}), resulting to the action determining the evolution of perturbations. However, since the involved expressions are very long and complicated, we follow the approximation made in Refs. \cite{Gubitosi:2012hu, Bloomfield:2012ff}. To be specific, we focus on modes that are deep inside the horizon, $k^2/a^2\gg H^2$, and then impose the quasi-static approximation to discard the time derivatives. This approximation has been widely used in similar investigations (see \cite{Peirone:2017ywi} for a recent discussion). In the present work, in order to examine the propagating properties clearly, we perform the above approximation in two steps. Firstly, we only retain the kinetic terms, and according to \eqref{g}, \eqref{Stueckelberg trick 1} and \eqref{Stueckelberg trick 2}, the expression of the $T^0$-term can be derived as follows
\begin{align}
\label{T0}
 \sqrt{-g}\dot{f}_T T^0 = a^3\dot{f}_T \Big[ 3\dot{\pi}\dot{\psi} +a^{-2}\partial_i\pi \big( \partial_i\phi-2\partial_i\psi \big) \Big] + 2a^2\dot{f}_T\partial_i\psi\partial_i\chi +2\dot{f}_T a^2H\partial_i\pi\partial_i\chi ~.
\end{align}
According to \cite{Gubitosi:2012hu}, one can obtain the expression for the $R$ term as
\begin{align}
\label{R}
 \sqrt{-g}f_T R = 2a f_T \big( \partial_i\psi\partial_i\psi-2\partial_i\phi\partial_i\psi \big) -2a\dot{f}_T\partial_i\pi \big( 2\partial_i\psi-\partial_i\phi \big) -6a^3(f_T\dot{\psi}^2\!-\!\dot{f}_T \dot{\pi}\dot{\psi}) ~.
\end{align}

It is worth mentioning that the remaining terms do not contribute to our calculations for the moment. Inserting the above expressions into (\ref{action}), we obtain
\begin{align}
 S = \int d^4 x\frac{M_P^2}{2} \left(-2af_T\partial_i\psi\partial_i\psi + 6a^3f_T\dot{\psi}^2 + 4af_T\partial_i\phi\partial_i\psi 
+ 4a^2\dot{f}_T\partial_i\psi\partial_i\chi + 4\dot{f}_Ta^2H\partial_i\pi\partial_i\chi \right)~.
\end{align}
By going to Fourier space and diagonalizing the kinetic matrix corresponding to the above expression, we obtain the dispersion relation $k^8=0$, which means there are no propagating scalar degrees of freedom in $f(T)$ gravity. This result is consistent with 
\cite{Izumi:2012qj}, where they find that the only propagating scalar degree of freedom is introduced by the quintessence field. However, we emphasize that our current analysis only reveals the dynamics of the propagating modes in a highly symmetric background of FRW universe. It does not reveal the full features on degrees of freedom of $f(T)$ theory, and for this topic we refer readers to Ref.~\cite{Ferraro:2018tpu} for detailed discussions.

Finally after discarding the time derivatives of the fields, we obtain
\begin{align}
\label{final action}
 S = \int d^4 x \Big[ & \frac{M_P^2}{2} \left( -2af_T\partial_i\psi\partial_i\psi + 4af_T\partial_i\phi\partial_i\psi + 4a^2 \dot{f}_T \partial_i\psi \partial_i\chi + 4\dot{f}_T a^2 H \partial_i\pi \partial_i\chi \right) \nonumber\\
 & +a^3M^2\pi^2-a^3\phi\delta\rho_m \Big] ~.
\end{align}
Note that in the above action we have added two extra terms. The first is the standard non-relativistic matter coupling to   gravity $(1/2)T^{00}\delta g_{00}$. The other is the mass term of the Goldstone $\pi$ denoted with $M$, which will help us to investigate 
the effect of the scalar $\chi$ more transparently. We derive its explicit expression as follows.

According to (\ref{actionfin}) and (\ref{Stueckelberg trick 1}), the mass term of the Goldstone mode $\pi$ can be produced by using the St\"uckelberg trick, which reads
\begin{align}
 \sqrt{-g} \frac{M_P^2}{2}\Psi(t)R \longrightarrow & a^3\frac{M_P^2}{4}\ddot{\Psi} R^{(0)}\pi^2 ~, \\
 -\Lambda(t) \sqrt{-g} \longrightarrow & -\frac{1}{2}\ddot{\Lambda}(t)a^3\pi^2 ~, \\
 -b(t)\sqrt{-g}g^{00} \longrightarrow & \frac{1}{2}\ddot{b}(t)a^3\pi^2 ~, \\
 \sqrt{-g}\frac{M^2_P}{2}d(t)T^0 \longrightarrow & -\frac{3}{4}Ha^3 M^2_P\ddot{d}\pi^2 ~.
\end{align}
Then, using also (\ref{cc}) and (\ref{ll}), we can extract the mass term as
\begin{align}  	
 M^2 = 3\dot{H}b -\frac{1}{4}M^2_P\dot{R}^{(0)}\dot{\Psi} +\frac{9}{4}M^2_PH\dot{H}d + \frac{3}{4} M^2_P\dot{H}\dot{d} ~.
\end{align}
In the case of $f(T)$ gravity, i.e. under the identifications provided in \eqref{fTidentification}, the above mass term becomes
\begin{align}
\label{mass}
 \frac{M^2}{M^2_P} = \frac{1}{4}\dot{R}^{(0)}\dot{f}_T +\frac{9}{2}H\dot{H}\dot{f}_T + \frac{3}{2}\dot{H}\ddot{f}_T ~.
\end{align}

We now proceed to the variation of the action (\ref{final action}) with respect to $\psi$, $\pi$, $\chi$ and $\phi$, which result to the following Poisson equation
\begin{align}
\label{G}
 k^2\phi=\frac{a^2}{2M^2_Pf_T}\left(1-\frac{a^2M^2}{M^2_Pf_TH^2k^2}\right)\delta\rho_m ~.
\end{align}
Furthermore, the post-Newtonian parameter $\gamma$ writes as
\begin{align}
 \gamma=\frac{\psi}{\phi}=1+\frac{a^2M^2}{f_TH^2k^2M^2_P-a^2M^2} ~.
\end{align}
Note that for TEGR we have $f_T= -1$ and $M^2 \longrightarrow 0$, and thus the above two expressions can recover the standard results of Newtonian gravity.

In order to examine the effect of additional scalar $\chi$, we let $\chi$ vanish in the action (\ref{final action}) and we repeat the calculations, finding that the second term inside the parentheses of expression (\ref{G}) vanishes, namely $k^2 \phi = a^2/2 M^2_P f_T$, implying that the post-Newtonian parameter $\gamma$ is exactly equal to 1. Hence, it is the additional scalar $\chi$ that makes the results to have the dependence on the mass-term of $\pi$. Nevertheless, note that according to (\ref{mass}) we deduce that $M^2/M^2_P\sim H^4$, while under our approximation we require $k^2/a^2\gg H^2$, which implies that the mass term is almost negligible. In summary, we conclude that the effect of the additional scalar $\chi$ is very small under the Newtonian limit, and thus we expect its effect to become apparent at the scale $k^2/a^2\sim H^2$.

Let us comment here that according to the above expressions we do verify that in $f(T)$ gravity the effective gravitational constant is $G_{eff}\simeq G/f_T$, which is a known result \cite{Zheng:2010am, Nunes:2018xbm}. Moreover, we do verify the known result that the effect of the scalar perturbation $\chi$ on the large-scale structure, namely the deviation from $\Lambda$CDM paradigm, becomes more important on large scales \cite{Dent:2011zz, Nesseris:2013jea}. Nevertheless, through the analysis of the present 
manuscript, and incorporating the second-order perturbation for the tetrads from the perspective of EFT, one may have a new window to explore torsional gravity and its cosmological implications.

\section{Applications in specific $f(T)$ cosmological models}
\label{applicspecficmods}

In this section we use the information obtained above in order to explicitly investigate two specific classes of viable $f(T)$ cosmological models, which are known to pass the current observational requirements \cite{Nesseris:2013jea, Nunes:2016qyp}. In particular, we focus on the possible observational signatures of $f(T)$ gravity on the post-Newtonian experiments in near future.

\subsection{The power-law model}

As a first model we consider the power-law model \cite{Bengochea:2008gz} which reads
\begin{align}
\label{powermodd}
 f(T) = -T +\alpha T^p ~,
\end{align}
with $\alpha$ and $p$ the model parameters. Knowing the present values of the matter density parameter $\Omega_{m0}$ and of the Hubble function $H_0$, we can express $\alpha$ as
\begin{align}
\label{eq:alpha01}
 \alpha = (6H^2_0)^{1-p} \frac{1-\Omega_{m0}}{2p-1} ~,
\end{align}
where we have assumed that at present the radiation density parameter is negligible, $\Omega_{r0}=0$, for simplicity. Note that in this type of cosmological models $\Lambda$CDM cosmology can be recovered when $p=0$.

Starting from the Friedmann equations (\ref{f11}) and (\ref{f22}), and using as the independent variable the redshift $z=\frac{a_0}{a}-1$ (we set the current scale factor $a_0$ to one for simplicity) we easily obtain
\begin{align}
\label{eq:fired01}
 y \equiv \frac{H^2}{H_0^2} = \Omega_{m0}(1+z)^3 +(1-\Omega_{m0})y^p ~.
\end{align}
Moreover, using (\ref{rhodein22}) and (\ref{pdein22}) we can define the effective dark energy equation-of-state parameter as follows,
\begin{align}
\label{wDEMod1}
 \omega_{DE}^{\text{eff}}(z) \equiv \frac{p_{DE}^{\text{eff}}}{\rho_{DE}^{\text{eff}}} = \frac{(p-1) y}{[p\Omega _{m0} (1+z)^3+(p-1)y ]} 
 \left[ -1 + \frac{2p (1+z)^3 \Omega_{m0}}{y-p(1-\Omega_{m0}) y^p} \right] ~.
\end{align}

One may use cosmological observations of the background level to extract the constraints on the model parameter $p$ \cite{Nesseris:2013jea, Nunes:2016qyp, Xu:2018npu, Nunes:2018xbm, Capozziello:2015rda}. However, in the present work we are interested in examining whether it is possible to constrain $p$ using the post-Newtonian observations and the perturbation analysis of the previous section. In particular, we introduce the parameter
\begin{eqnarray}
 \mu\equiv\frac{1}{f_T} ~,
\end{eqnarray}
which can quantify the deviation from general relativity due to the non-trivial $f(T)$ form (in the conventions we use in this work the general relativity value is $-1$). Taking into account the perturbation analysis of the previous section, and specifically the Poisson equation (\ref{G}), we can express $\mu(z)$ as
\begin{eqnarray}
 \mu(z)=\frac{2M^2_Pk^2\phi (1+z)^2}{\delta\rho_m} ~,
\end{eqnarray}
under the Newtonian limit, which is therefore related to cosmological observations.

\begin{figure}[htbp]
\includegraphics[width=0.45\textwidth]{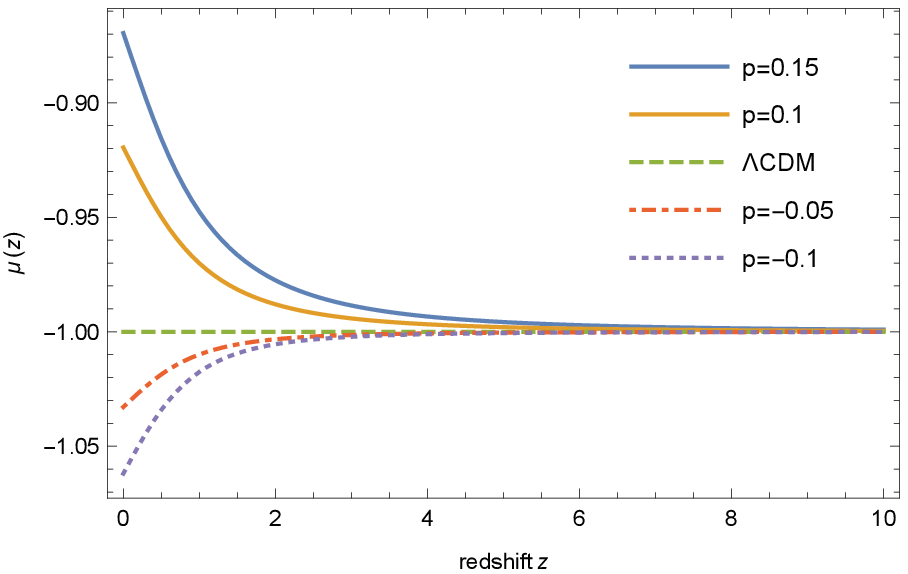}
\includegraphics[width=0.45\textwidth]{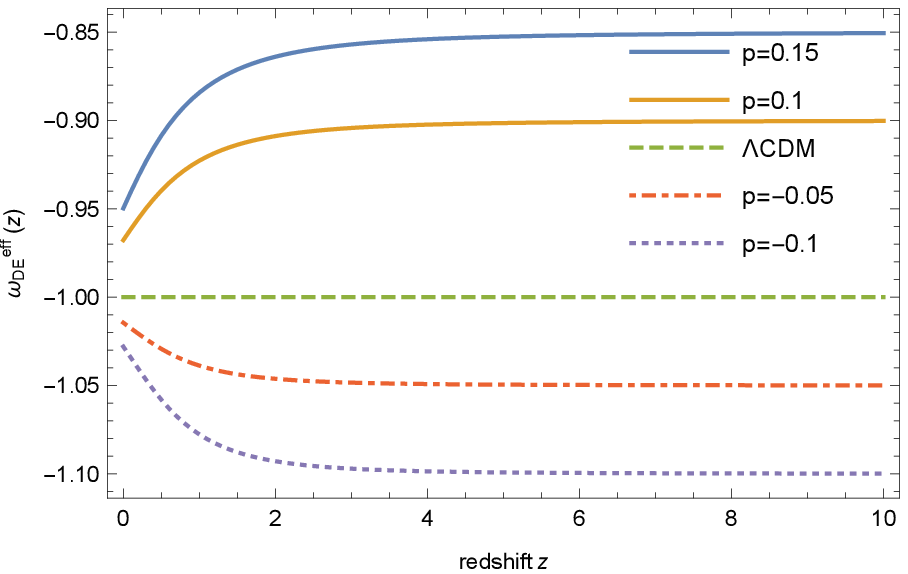}
\caption{
{\it{
Left graph: The evolution of the deviation quantity $\mu=1/f_T$ as a function of the redshift $z$, for the case of the power-law $f(T)$ model (\ref{powermodd}), for various $p$ values. We have imposed the condition that $\Omega_{m0}=0.3$ in the numerical estimates. Right graph: The corresponding evolution of the effective dark energy equation-of-state parameter $\omega_{DE}^{\text{eff}}(z)$.
}}}
\label{figure mu}
\end{figure}

In the left graph of Fig. \ref{figure mu} we depict $\mu(z)$ as a function of the redshift for various values of the parameter $p$. As we can see, different values of $p$ correspond to different dynamical behaviors of the gravitational coupling coefficient at low 
redshift. Hence, a possible precise measurement of $\mu$ could be used to constrain $p$. Finally, for completeness, in the right graph of Fig. \ref{figure mu} we present the corresponding evolutions of the effective dark energy equation-of-state parameter 
$\omega_{DE}^{\text{eff}}(z)$ according to (\ref{wDEMod1}).

\subsection{The exponential model}

As a second specific example we will consider the exponential model 
\cite{Nesseris:2013jea}:
\begin{align}
\label{eq:model2}
 f(T) = -T +\beta T_0(1-e^{-q T/T_0}) ~,
\end{align}
with $q$ the dimensionless model parameter, while for the other parameter $\beta$ we have
\begin{align}
\label{eq:beta}
 \beta= {1-\Omega_{m0}\over -1+(1+2q)e^{-q}} ~,
\end{align}
with $T_0=6H_0^2$ the present value of $T$. This type of models can reduce back to standard $\Lambda$CDM cosmology under the limit of $q\rightarrow +\infty$.

For this type of models, the Friedmann equations (\ref{f11}) and (\ref{f22}) yield the following equation
\begin{align}
\label{eq:fired02}
 y \equiv \frac{H^2}{H_0^2} = \frac{e^{q-q y} (1+2qy-e^{q y}) (1-\Omega _{m0})}{1+2q-e^q} + (1+z)^3 \Omega_{m0} ~.
\end{align}
Additionally, expressions (\ref{rhodein22}) and (\ref{pdein22}) lead to an effective dark energy equation-of-state parameter of the form
\begin{align}
\label{wDEMod2}
 \omega_{DE}^{\text{eff}} = -1 + \frac{(1-2 q y) (\beta qe^{-q y} -1 )}{[1-e^{q y}+q(y+(1+z)^3 \Omega _{m0})]}
 \left[ \frac{q e^{q y} (1+z)^3 \Omega_{m0}}{\beta q(1-2 q y)-e^{q y}} \right] ~.
\end{align}

\begin{figure}[htbp]
\includegraphics[width=0.45\textwidth]{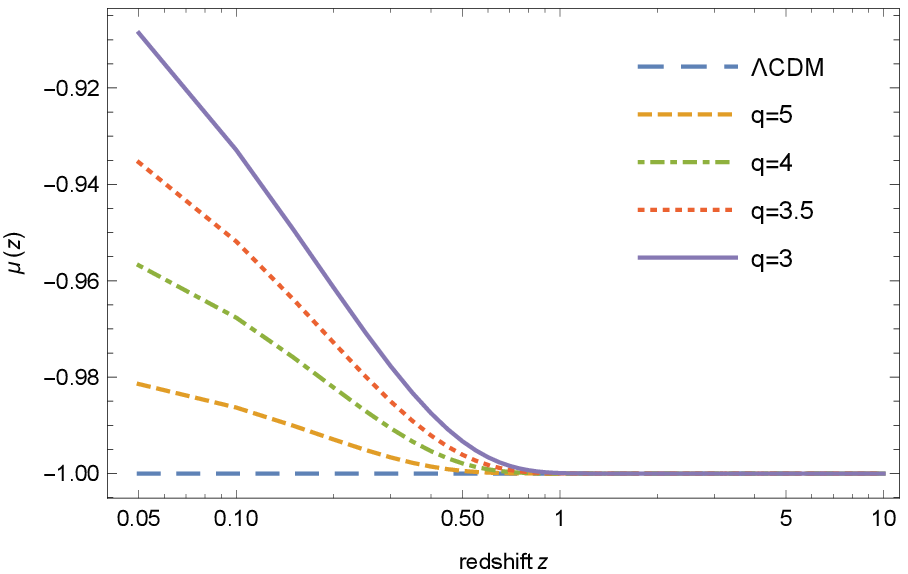}
\includegraphics[width=0.45\textwidth]{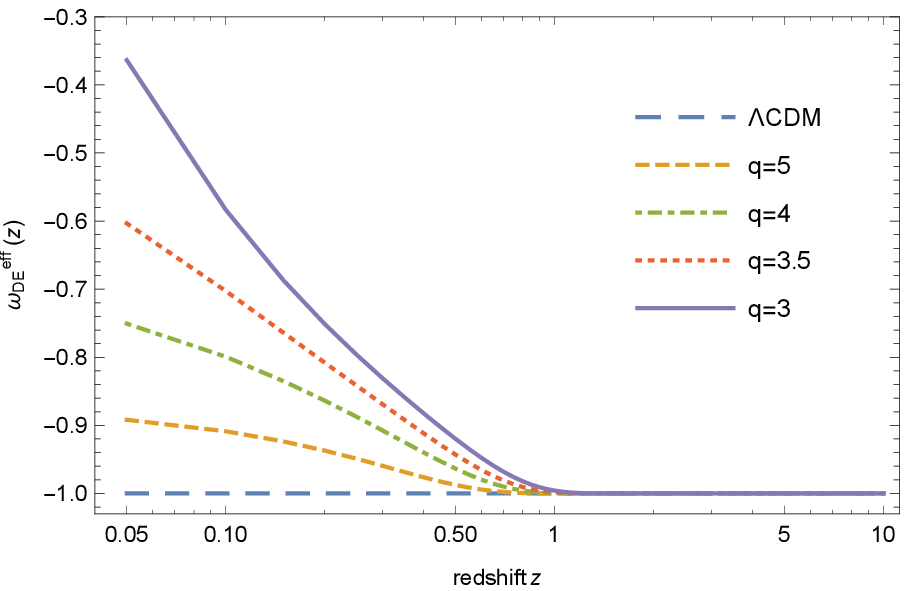}
\caption{
{\it{
Left graph: The evolution of the deviation quantity $\mu=1/f_T$ as a function of the redshift $z$, for the case of the exponential $f(T)$ model (\ref{eq:model2}), for various $q$ values. We have imposed the condition that $\Omega_{m0}=0.3$ in the numerical 
estimates. Right graph: The corresponding evolution of the effective dark energy equation-of-state parameter $\omega_{DE}^{\text{eff}}(z)$.
}}}
\label{figure mu-2}
\end{figure}

In the left graph of Fig. \ref{figure mu-2} we present the evolution of $\mu\equiv1/f_T$ as a function of the redshift for various values of the parameter $q$. As we observe, and as expected, as $q$ gradually decreases from $+\infty$ the exponential model 
(\ref{eq:model2}) gradually deviates from the $\Lambda$CDM cosmology. Similarly to the previous model, a possible precise measurement of $\mu$ could be used to constrain $q$. Additionally, we mention that since the behavior of $\mu(z)$ changes for different $f(T)$ models, the measured $\mu(z)$ could help to break the degeneracy between them. Finally, in the right graph of Fig. \ref{figure mu-2} we present the corresponding behavior of the effective dark energy equation-of-state parameter $\omega_{DE}^{\text{eff}}(z)$ according to (\ref{wDEMod2}). As we can see, its present value algebraically increases for decreasing $q$.

\section{Conclusions}	
\label{Conclusion}

In the present work we developed the EFT approach to the theory of torsional gravity. The EFT formalism is very efficient, since it allows for the systematic investigation of the background and perturbation levels of a theory separately, without the need to incorporate all the details of the underlying fundamental theory.

In order to perform the realization of the EFT approach to torsional gravity under the teleparallel condition, both the background and the perturbation part of the usual EFT approach (i.e the EFT approach to curvature-based gravity) needed to be generalized by including several appropriate terms. For the background part we needed to add terms of $T^0$, i.e of the contracted torsion tensor, which is actually the boundary term that relates the torsion and curvature scalars, reflecting in the EFT language. On the other hand, for the perturbation part, apart from the known pure curvature terms of the usual EFT approach, we needed to include pure torsion perturbative terms and mixed perturbative terms of torsion and curvature. In summary, we were able to extract the EFT action, at both background and perturbation levels, of all torsional modified gravity. Knowing the EFT action of general torsional modified gravity allows to proceed to the investigation of sub-classes of torsional gravity by suitably choosing the various action terms. We focused on $f(T)$ gravity in detail, nevertheless we also provided the method to study other sub-classes, such as scalar-torsion theories or $f(T,R)$ gravity.

We then proceeded to the cosmological application of the EFT approach to $f(T)$ gravity. In particular, we were able to investigate the perturbations up to second order, although the study of higher orders is straightforward. Taking advantage of the fact that the 
background evolution and the perturbations can be separated under the EFT formalism, we studied the scalar perturbations of $f(T)$ theory, focusing on the effect of the scalar mode $\chi$. Under the quasi-static approximation and requiring that the modes are deep inside the horizon, we derived the expressions of the Newtonian constant $G$ and the post Newtonian parameter $\gamma$. We found that the $f(T)$ scalar perturbation effect is to make the results to depend on the mass term of the Goldstone mode $\pi$. Nevertheless, since in our approximation the mass term is small, the above effect is suppressed and we expect it to be manifested at the horizon crossing scale.

Finally, we applied this procedure to two specific and viable $f(T)$ models, namely the power-law and the exponential ones, introducing the new parameter $\mu(z)$ that quantifies the deviation from general relativity and depends on the model parameter. On the other hand, according to our analysis $\mu(z)$ was expressed in terms of the scalar perturbation mode, and thus we were able to provide plots of its behavior for different values of the model parameters. Hence, a possible precise measurement of $\mu(z)$ could be used as an alternative way to impose constraints on $f(T)$ gravity and break possible degeneracies 
between different $f(T)$ models.

\begin{acknowledgments}
We are grateful to Y.-S. Piao, L. Xue, Y. Zhang and W. Zhao for valuable comments. YFC and CL are supported in part by the Chinese National Youth Thousand Talents Program, by the NSFC (Nos. 11722327, 11653002, 11421303), by the CAST Young Elite Scientists Sponsorship (2016QNRC001), and by the Fundamental Research Funds for the Central Universities. YFC is grateful to ENS for his hospitality at the National Technical University of Athens, Greece during the preparation of this work. YC would like to thank Burt Ovrut and Rehan Deen for hospitalities during his visit at the University of Pennsylvania. YC is supported in part by the UCAS Joint PhD Training Program. ENS is supported in part by the COST Action ``Cosmology and Astrophysics Network for Theoretical Advances and Training Actions'' and by COST (European Cooperation in Science and Technology). Part of numerics are operated on the computer cluster LINDA in the particle cosmology group at USTC.
\end{acknowledgments}

%-----------------------------------

%---------------------------

\end{document}